\documentclass[10pt, conference, compsocconf]{article}
\usepackage[utf8]{inputenc}
\usepackage[english]{babel}
\usepackage{amsmath}
\usepackage{amsfonts}
\usepackage{amssymb}
\usepackage{graphicx}
\usepackage{url}
\usepackage{cite}

\usepackage{color}
\usepackage[tight,footnotesize]{subfigure}

\begin{document}

\title{Steganalyzer performances in operational contexts}

\author{Yousra Ahmed Fadil, Jean-Fran\c cois Couchot, Rapha\"{e}l Couturier,\\ and Christophe Guyeux}

\maketitle

\begin{abstract}
Steganography and steganalysis are two important branches of the information hiding field of research. 
Steganography methods consist in hiding  information in such a way that the secret message is undetectable for the uninitiated. Steganalyzis encompasses all the techniques that attempt to detect the presence of such hidden information. 
This latter is usually designed by making classifiers able to separate innocent images from steganographied ones according to their differences on well-selected features. We wonder, in this article whether it is possible to construct a kind of universal steganalyzer without any knowledge regarding the steganographier side. The effects on the classification score of a modification of either parameters or methods between the learning and testing stages are then evaluated, while the possibility to improve the separation score by merging many methods during learning stage is deeper investigated.
%\begin{color}{red}The steganalyzis for images is a more hard and challenging task because the data embedding  take part different parameters such as image formats,  embedding algorithms.  In this paper, the state of the art steganalyzer system was tested against changing the popular steganography methods in training and testing set. Also, the effect of using different payloads was noticed. And as a result of an existence the noise in images and there are many types of filters to deal with these type of noises, we try to see the effect of some type of smoothing method on the steganalyzer. The results of these experiments were different depending on the strength of the effect of these parameters on the steganalyzis system. \end{color}
\end{abstract}

\section{Introduction}
%\input{introduction}

Information hiding is a recent
computer science security field that focuses on the capability to hide hidden messages in digital media like pictures or movies~\cite{bgh13:ip,bcfg+13:ip}. This 
discipline encompasses the design of algorithms, called steganographiers, that aim at discreetly inserting secret messages into innocent like 
cover media, the output being called stego-content. Conversely, steganalysers are
tools that aim at detecting the possible presence of a secret message in a given document. Available steganalysers are mainly based on ensemble classifiers that have learned to separate between real natural images and stego contents.

Steganalyzers of the literature are usually evaluated as follows. A state-of-the-art steganographier $s$ is firstly chosen, while the BOSS images~\cite{BOSS} are separated in two sets, half of each two parts being steganographied using $s$. Then the first set is used during the learning stage, while the steganalysis method is evaluated using the second set.
Such an evaluation corresponds to the particular situation where the warden Eve (the steganalyzer) has the knowledge of which steganographier has been used, with which parameters (embedding payload, etc.) In this article, we will investigate a more realistic scenario where Eve only knows that images contain secret messages: she does not know which steganographic algorithm has been used, and the game consists of separating well original from stego contents.
More precisely, in this research work, we show what happens when the learning stage has been realized with a wrong steganographier, and we ask whether it is useful to use more than one steganographier during the learning stage to face this problem.

The remainder of this article is constituted as follows. In the next section, we briefly recall the functioning of state-of-the-art steganographiers and steganalyzers studied in this article. 
Then, in Section~\ref{sec:Eve}, we first investigate the effect of a wrong assumption on the steganographier during the learning stage. 
In Section~\ref{sec:mixingStego}, we 
wonder whether it is possible to solve
this problem by mixing more than one 
steganographier during the learning 
stage, in order to design a kind of
universal detector. Errors on payload
assumption are then discussed in
Section~\ref{sec:payload}. All these situations are merged in Section~\ref{sec:operational}, leading to what can be expected for operational contexts.
%This article ends with a conclusion section, in which the study is summarized and intended future %work is outlined.

\section{State of the art}
Let us now recall some famous algorithms that will be investigated in this article. The first next paragraph focuses on steganographiers while the second one is about steganalyzers. Readers wanting more details about these schemes are referred to the provided citations.

In F5 algorithm~\cite{f5}, the absolute value of some randomly selected DCT coefficients is decreased by one. However, to avoid errors during decoding, F5 algorithm skips over all the coefficients equal to +/-1 which is denoted as shrinkage.
The nsF5~\cite{nsF5} algorithm is introduced as a modified version of F5 by alleviating the shrinkage. HUGO~\cite{hugo1} method, for its part, focuses on so-called efficient SPAM features on spatial domain. Finally, in universal distortion function J-UNIWARD~\cite{J-uniward}, the embedding 
is performed on specific regions of the cover objects, more precisely on texture and noisy ones, where the distortion function is computed according to wavelet domain. This method avoid the embedding in clean edges and smooth regions.

%Hiding information within a given image alter some of its characteristics, and such an alteration may reveal the existence of a secret message. 

Ensemble classifiers have been proposed
as steganalysis systems in~\cite{ensemble}. They are
built by fusing decisions of an ensemble of simple base learners that are inexpensive to train, leading to a steganalyzer of low complexity.
To achieve this goal, these schemes explored several different possibilities for the base learners and fusion rules for designing the final classifier. This latter has been improved in 
%The steganalysis systems considered more robust when depend on the features extracted from the images. 
CC-PEV, whose functioning is detailed in~\cite{pevny2007merging}. %is extracted Markov and DCT Features. The two types of features are extracted from the DCT domain. 

\section{Training and testing stages use not the same steganographier}
%\section{Eve's ability to separate between natural and stego contents without steganographier knowledge}
\label{sec:Eve}
%\subsection{Presentation of experiments}

%\subsubsection{Training and testing stages do not share the same steganographier}
Let us first measure the effects of modifying the steganographic method between the training and the testing stage. %. 2,000 images have been used as natural images, and the embedded message is constituted by
To investigate this question, $2,000$ original JPEG images have been used in our experiments.
They are taken from the BOSS contest~\cite{BOSS},
their size is equal to $512 \times 512$, and they have been converted to JPEG. %. These images are converted to jpeg images using Matlab imwrite command. 
For stego images, an embedding payload of 0.1 is used. The method used for extracting the features from the images is CC-PEV.  %Two methods are used to embed the secret message nsF5~\cite{nsF5}, J-UNIWARD~\cite{J-uniward}. 
The same ensemble classifier has been used both in the training and in the testing stage,
namely the one of~\cite{ensemble}.

\begin{figure}[!h]
  %\vspace{-0.2cm}
\centering
\subfigure[nsF5 for training, and J-UNIWARD for   
          testing]{\includegraphics[width=7cm,height=5cm]{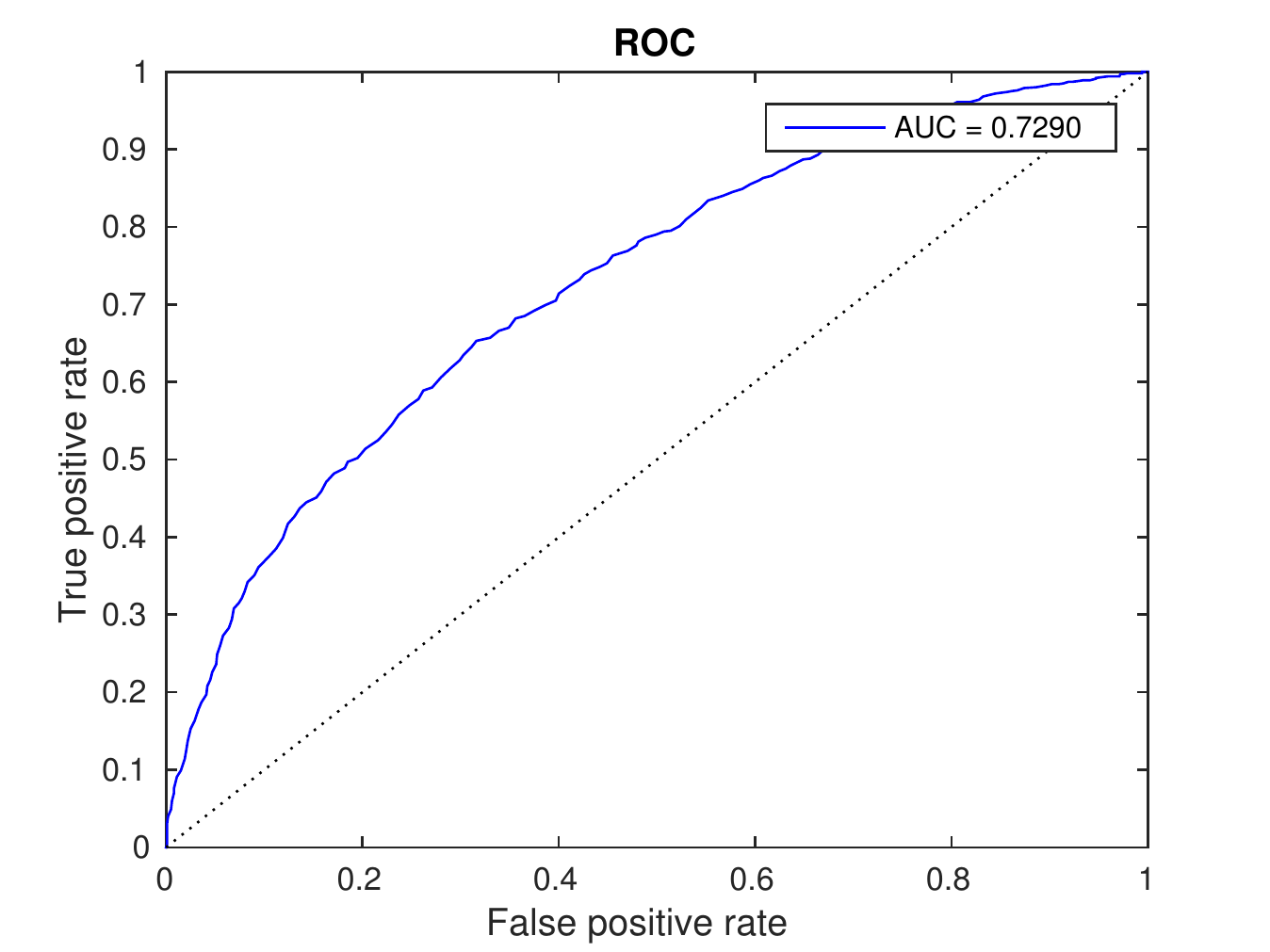}}
\subfigure[The converse]{\includegraphics[width=7cm,height=5cm]{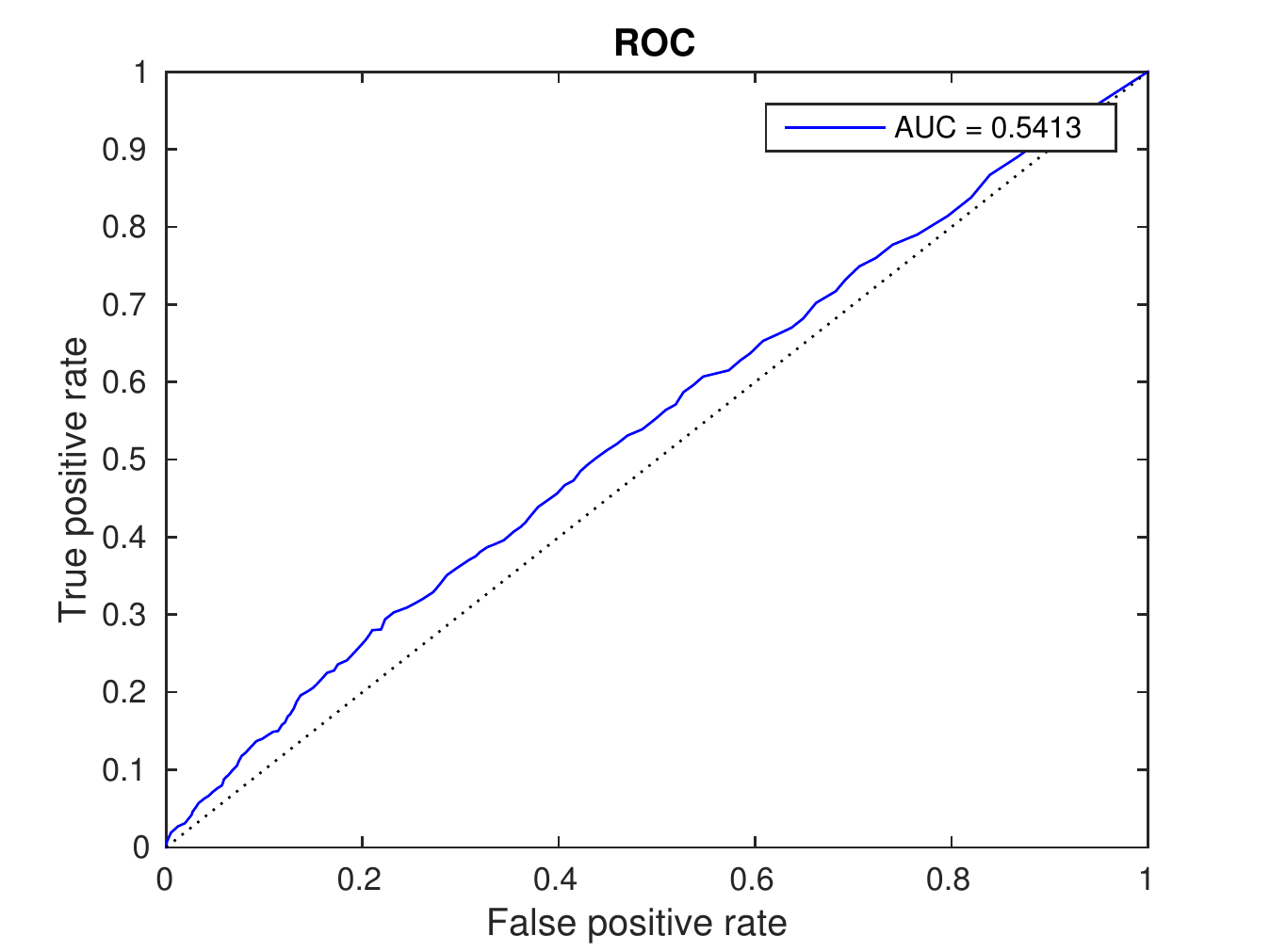}}
 \caption{The steganographier used during the training stage is not the good one.}
\label{fig:f1}
\end{figure}

In the first experiment, the ensemble classifier is trained using 50\% of the natural images and  50\% of the same images steganographied by nsF5, while it is tested using the  same rate of natural and J-UNIWARD images. Conversely, in the second set of experiments, J-UNIWARD is used during the training stage and nsF5 during the testing one. Obtained results are presented in Figure~\ref{fig:f1}. It can be seen that the presence of J-UNIWARD hidden messages is more or less detected when the steganalyzer has been trained by using nsF5. Conversely, 
the detection of nsF5 is impossible when
learning with J-UNIWARD. This asymmetric behavior may be explained by the fact that the use of nsF5 affects more the general aspect of the embedding image compared to J-UNIWARD. So, the ensemble classifier can learn more from the former than from the latter, and its classification is thus more efficient and trustworthy. This result has been obtained again when considering all other possible combinations, see Table~\ref{tab:mixingStego}: the only 
acceptable performances are obtained when nsF5 is used during the training stage.

\begin{table}[h]
\centering
\begin{tabular}{|c|c|c|c|}
\hline
Learning stage & Testing stage & A.U.C.& A.T.E \\
\hline
nsF5 & J-UNIWARD & 0.7377& 0.3569\\
J-UNIWARD & nsF5 & 0.5426& 0.4675\\
nsF5 & HUGO & 0.7523&0.3345\\
HUGO & nsF5 & 0.5371&0.4737\\
J-UNIWARD & HUGO & 0.5122& 0.4912\\
HUGO & J-UNIWARD &  0.5077& 0.4915\\
\hline
\end{tabular}
\caption{Errors when choosing the learning steganographier}
\label{tab:mixingStego}
\end{table}

\section{Trying to improve steganalyzer score by mixing learning steganographiers}
\label{sec:mixingStego}

In this new scenario, we wonder whether the steganalysis performance can be improved by using more than one steganographier during the learning stage: if two or three steganographiers are suspected by Eve, can she use such a suspicion to produce a more accurate steganalyzer ? Or, to say this differently, is it possible to create a kind of universal steganalyzer by using a large set of steganographiers during the learning stage ? 

\begin{table}[h]
\centering
\begin{tabular}{|c|c|c|c|}
\hline
 Learning stage & Testing stage & A.U.C& A.T.E \\
 \hline
% Nat hugo juniward nsF5
(50, 25, 0, 25) & (50, 0, 0, 50) &0.6899& 0.3584\\
(50, 25, 0, 25) & (50, 50, 0, 0) &0.6097& 0.4269\\ 
\hline
(50, 0, 25, 25) & (50, 0, 50, 0) & 0.6133&0.4284\\ 
(50, 0, 25, 25) & (50, 0, 0, 50) & 0.6914&0.3518\\ 
\hline
(50, 25, 25, 0) & (50, 50, 0, 0) &0.5104& 0.4920\\ 
(50, 25, 25, 0) & (50, 0, 50, 0) &0.5208& 0.4855\\ 
\hline
(50, 25, 25, 0) & (50, 0, 0, 50) &0.5415& 0.4692\\ 
(50, 25, 0, 25) & (50, 0, 50, 0) & 0.6039&0.4306\\ 
(50, 0, 25, 25) & (50, 50, 0, 0) &0.6158&0.4149\\ 
\hline
(50, 25, 25, 0)  &(50, 25, 0, 25) &0.5404&0.4718\\ 
(50, 25, 0, 25)  &(50, 25, 25, 0) &0.6072&0.4303\\ 
(50, 0, 25, 25)  &(50, 25, 25, 0) &0.6585&0.4199\\ 
\hline
(50, 25, 0, 25) &(50, 0, 50, 0) & 0.6095&0.4311\\ 
(50, 0, 25, 25) &(50, 50, 0, 0) & 0.6321&0.4155\\ 
\hline
\end{tabular}
\caption{Study of accuracy by mixing various steganographiers when training. Each tuple represents the respective percentage of natural images, HUGO, J-UNIWARD, and nsF5 stego-contents.}
\label{tab:my_label}
\end{table}

\begin{figure}[!h]
\centering
\subfigure[In the training stage: 1000 natural images, together with 500 nsF5 and 500 HUGO stego-contents. In the testing stage: 1000 natural and 1000 JUNIWARD.]{   \includegraphics[width=7cm,height=5cm]{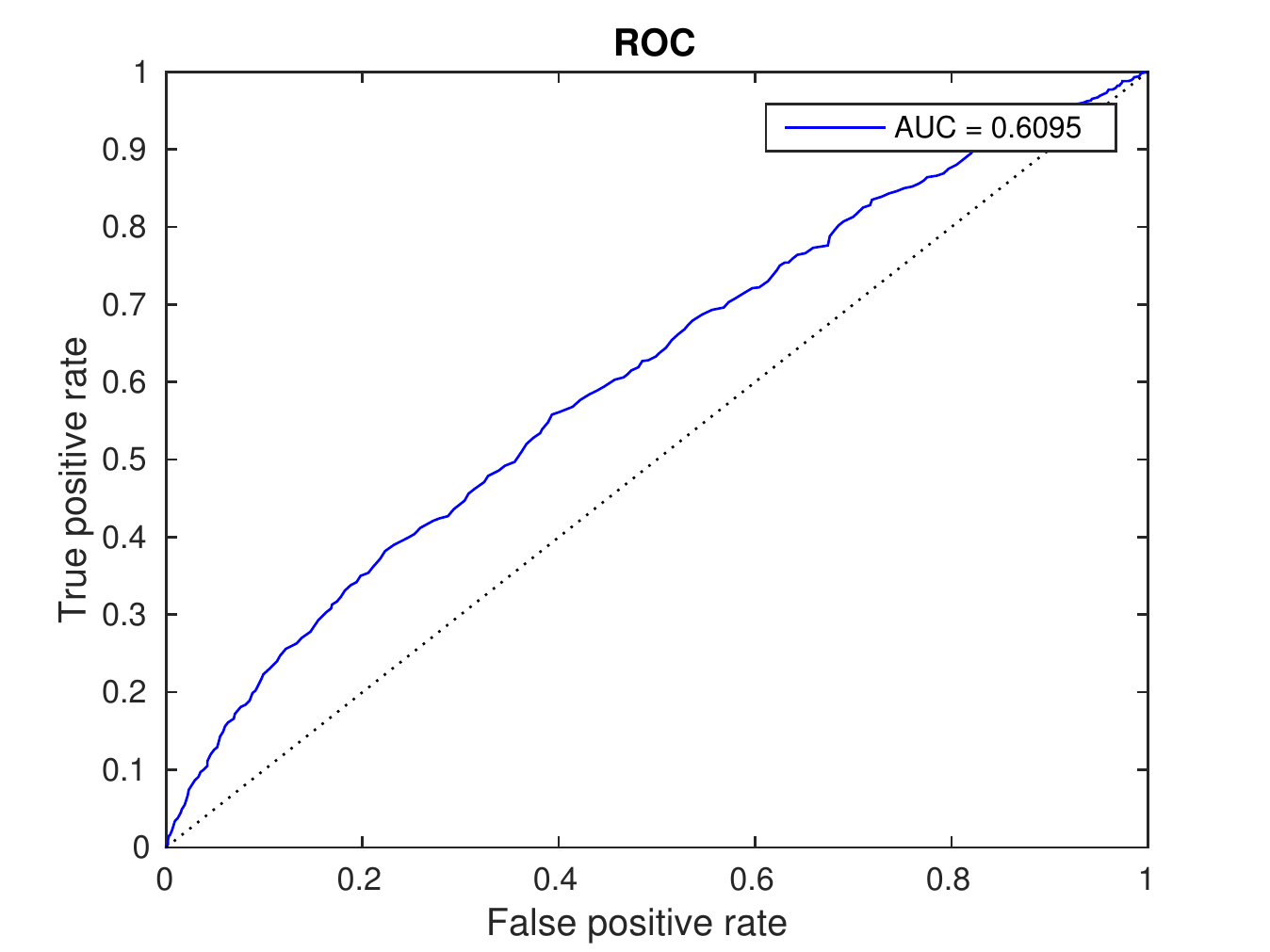}
}
\subfigure[In the training stage: 1500 natural + 1500 JUNIWARD, while in the testing set: 500 natural and 500 JUNIWARD]{   
   \includegraphics[width=7cm,height=5cm]{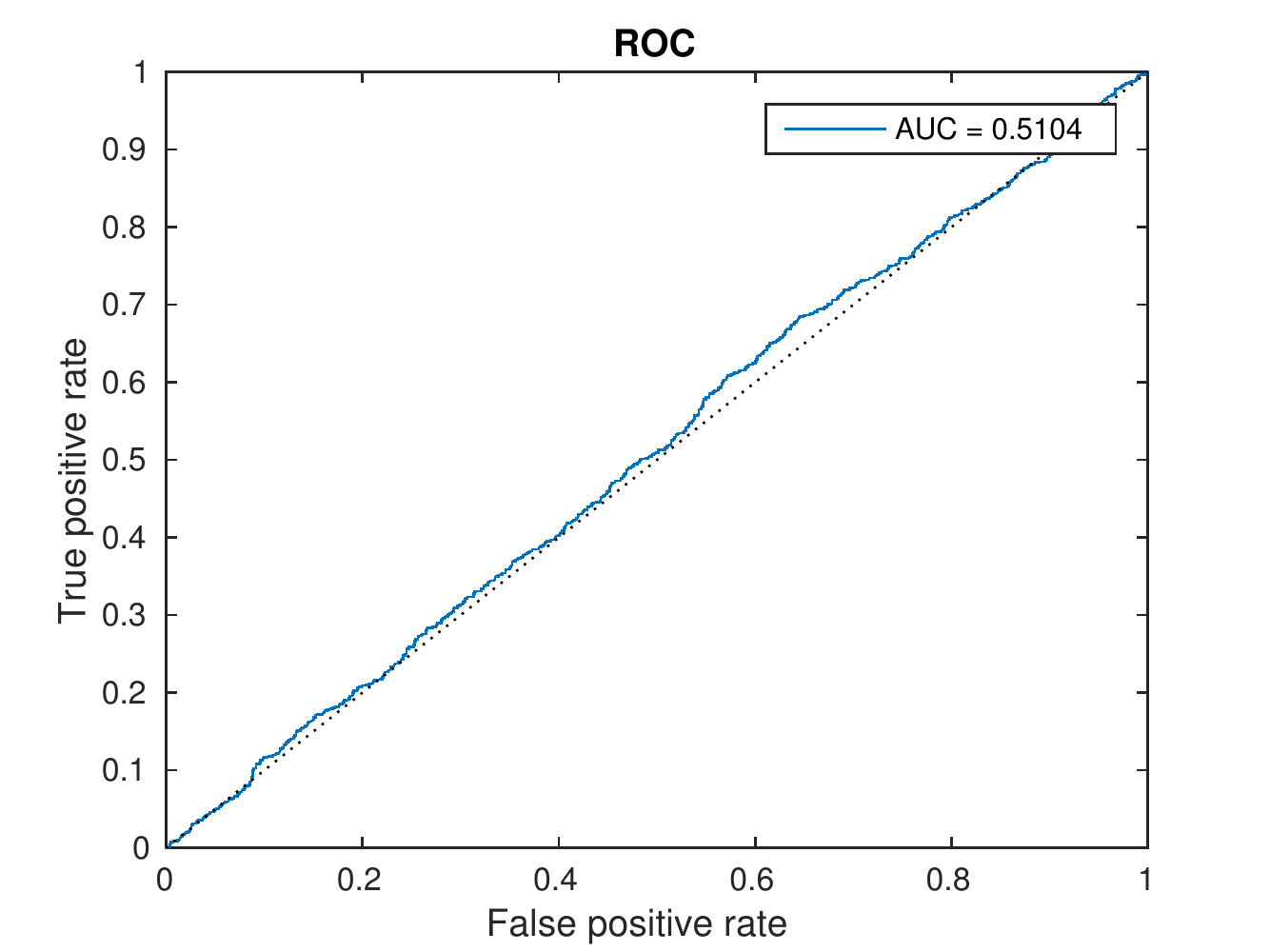}
}
  \caption{Mixing various steganographiers in the learning stage.}
  \label{fig:mix22}
 \end{figure}

Results of these experiments are given in Table~\ref{tab:my_label} and partially illustrated in receiver operating characteristic (ROC) curves of Figure~\ref{fig:mix22}. In this table, each row corresponds to an experiment where more than one steganographier has been used during the learning stage. Each tuple in this table gives the proportion of, respectively, natural images, HUGO, J-UNIWARD, and nsF5 stego-contents that has been used to constitute the set of 2,000 images, either during training or during testing stage. A payload of 0.1 has been used as previously. However, the area under the curve (AUC) obtained here never becomes larger than 0.7, while it was the case in Table~\ref{tab:mixingStego},  setting at naught the hope to constitute universal steganalyzer by mixing several tools when training.

\section{Uncertainty effects regarding payload}
\label{sec:payload}
The objective is now to emphasize the possible effects of payload ignorance on steganalyzer performances. Indeed, a large payload of 0.1 is always chosen for evaluating steganalyzers of the literature. By doing so, steganalyzer designers made strong assumptions that make life less complicated, and the game totally unfair in their own advantage. These two assumptions are that the steganographier will absurdly use a very large payload, and additionally this payload is known by the steganalyzer. Everything happens as if steganalyzer designers claim to be able to detect if a communication channel possibly contains stego images, while they finally answer to the challenge: ``knowing the set of images, the presence of hidden information, the steganographier, and the payload, can we separate with a good accuracy the natural from the stego images.''
On our side, we argue that it is not possible to expect exactly the payload value chosen by the steganographier in operational contexts.

\begin{figure}[!h]
  %\vspace{-0.2cm}
  \centering
   \includegraphics[width=7cm,height=1.8cm]{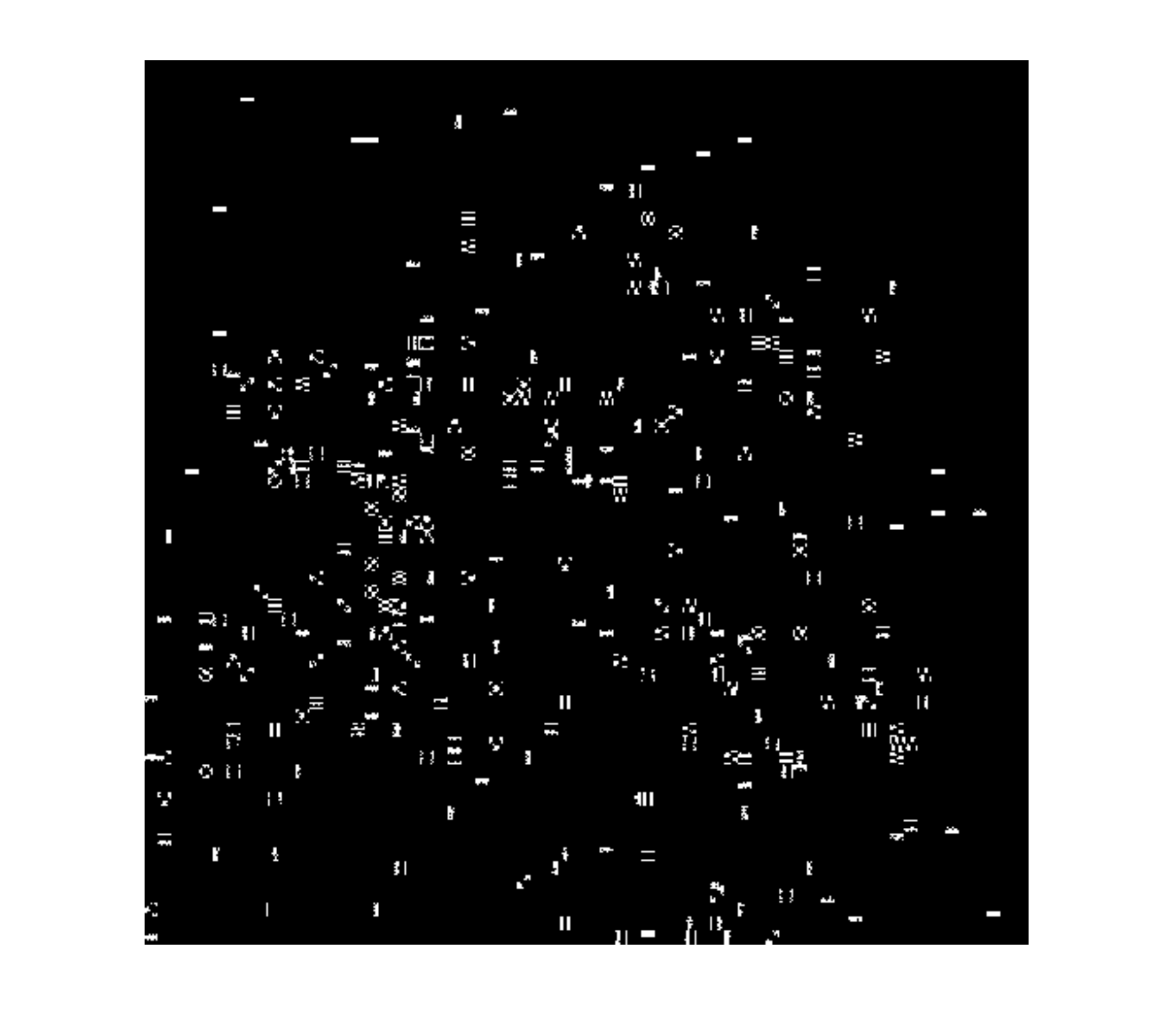}
   \includegraphics[width=7cm,height=1.8cm]{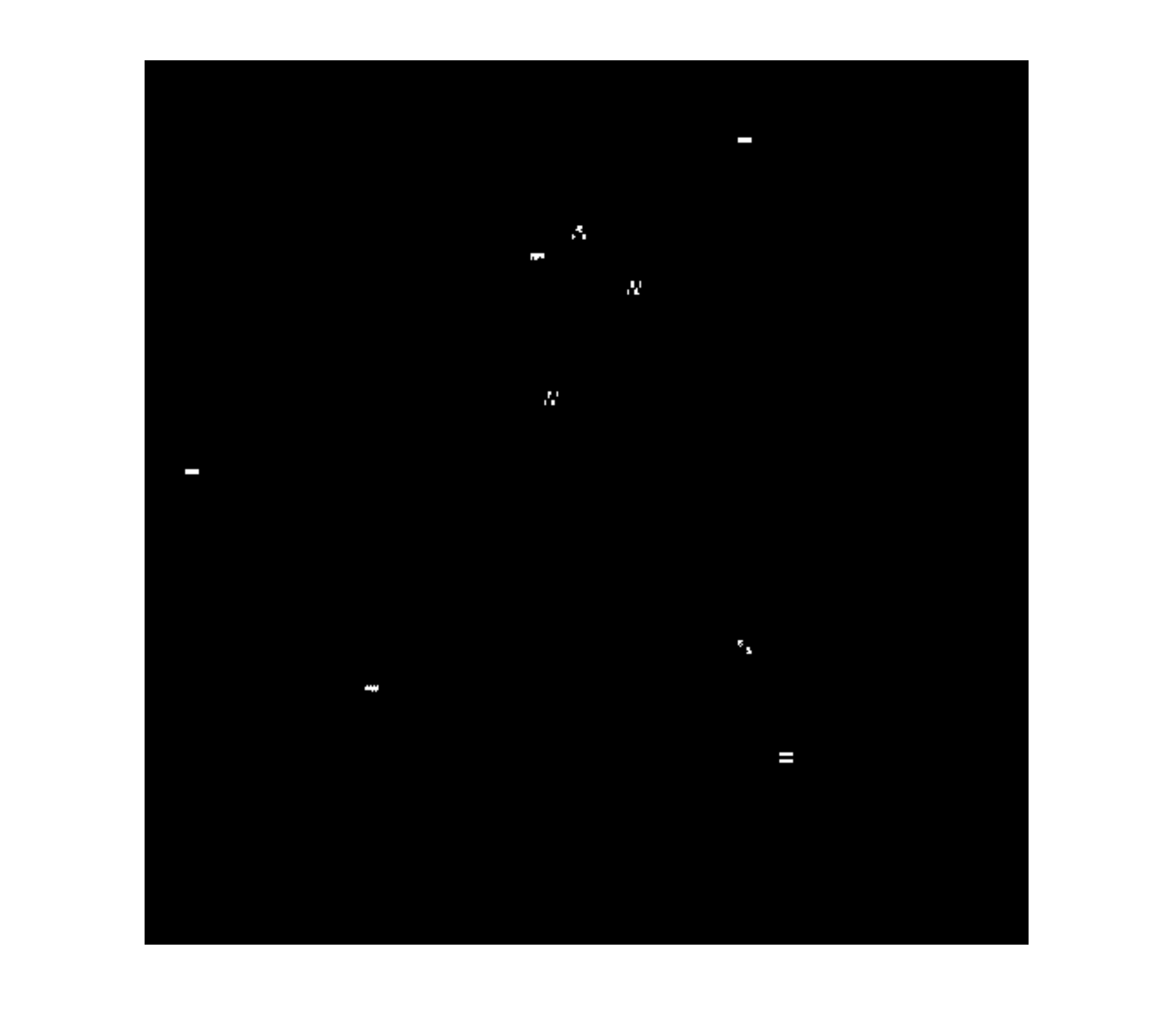}
  \caption{Differences between host content with nsF5 when payload is respectively equal to 0.1 or 0.005.}
  \label{fig:f4}
 \end{figure}

In this new run of tests, images are  steganographied by using respective payloads of 0.005, 0.05, and 0.1 (see  Figure~\ref{fig:f4} to understand the effects of such payloads on host contents).
CC-PEV features are used with ensemble classifier in both training and testing stages. %Different payloads are used, 0.005, 0.05, 0.1.
%In order to see the effect of using different payloads in training and testing stage, the following scenario was presented with different payloads and different embedding process work in DCT domain and special domain like nsF5, J-UNIWARD, HUGO. The AUC is  computed in each test and the results are presented in Table~\ref{tab:table1}
nsF5, J-UNIWARD, and HUGO have been successively tested using the 3 payloads listed above, to illustrate the effects of such an error for both spatial and frequency embedding. Obtained results are summarized in Table~\ref{tab:table1}.
As can be seen, the only situation where the separation is acceptable is the nsF5 one, and when training with a  large  payload that helps the ensemble classifier to learn the embedding effects.
 \begin{table}[h]
\centering
\begin{tabular}{|c|c|c|c|c|}
  \hline
              & Train        & Test       & A.U.C& A.T.E\\
  \hline
   nsF5       & 0.1    & 0.05  & 0.7855 &0.2854\\
  %\hline
              & 0.1    & 0.005 & 0.7723 &0.3195\\
  %\hline
              & 0.05   & 0.1   & 0.6656 &0.3933\\
  %\hline
              & 0.005  & 0.1   & 0.5408 &0.4717\\    
  \hline
   J-UNIWARD  & 0.1    & 0.05  &  0.5049 &0.4949\\
  %\hline
              & 0.1    & 0.005 &  0.5091&0.4955\\
  %\hline
              & 0.05   & 0.1   & 0.5087 &0.4958\\
  %\hline
              & 0.005  & 0.1   & 0.5035 &0.4980\\
  \hline           
  HUGO        & 0.1    & 0.05  &  0.5175&0.4898\\
  %\hline
              & 0.1    & 0.005 & 0.5161 &0.4885\\
  %\hline
              & 0.05   & 0.1   &  0.5182&0.4867\\
  %\hline
              & 0.005  & 0.1   &  0.5192&0.4853\\
  
  \hline
\end{tabular}
\caption{Payload error during training}
\label{tab:table1} 

\end{table}

\section{Operational contexts}
\label{sec:operational}

We now consider the most realistic scenario where the steganalyzer side only knows that one of the 3 most famous steganographier tools are used. But he is not sure about the chosen payload. Obtained results when mixing both the steganographier and its payload between training and testing stages have then been computed, and obtained results are summarized in Table~\ref{tab:table2}.

 \begin{table}[h]
 \centering
\begin{tabular}{|c|c|c|c|c|}
  \hline
                   & Train        & Test       & A.U.C&A.T.E\\
  \hline
  Train: nsF5 &     0.1  & 0.05  & 0.7388   &  0.3530 \\
  Test:J-UNIWARD&   0.1  & 0.005 & 0.7504   &  0.3325 \\
                &   0.05 & 0.1   & 0.5926   &  0.4482 \\
                &   0.005& 0.1   &  0.5057  &  0.4952 \\    
  \hline
  
   Train:nsF5       & 0.1         & 0.05  & 0.7489 &0.3320\\
    Test:HUGO
                   & 0.1          & 0.005 &  0.7554&0.3327\\
  
                   & 0.05         & 0.1   & 0.5791 &0.4378\\
  
                   & 0.005        & 0.1   & 0.5041 &0.4971\\
  \hline
  Train:J-UNIWARD   & 0.1         & 0.05  & 0.5140 &0.4905\\
  Test:HUGO
                   & 0.1          & 0.005 & 0.5119 &0.4916\\
  
                   & 0.05         & 0.1   & 0.5035 &0.4967\\
  
                   & 0.005        & 0.1   & 0.5020 &0.4997\\
  
  \hline
  
\end{tabular}
\caption{AUC scores in operational contexts}
\label{tab:table2}
\end{table}

As can be deduced from this table,
the classification is acceptable only
when the learning process has been realized with nsF5 and with a larger payload than the one that has been used during the tests. In this situation, 
it has been possible to separate, with a medium accuracy, images steganographied by either HUGO or J-UNIWARD. Remark that, obtained results are better than what has been found in Table~\ref{tab:table1}.

%\begin{color}{blue}
%From the previous results, it is observed that the problem with the small payloads. Where the classifier cannot discover the presence of the secret message. The performance of the classifier can be improved by using more that one payload in the learning stage. This can be seen in Figure~\ref{fig:f5}, in this scenario 3 payloads are used 0.1, 0.05, 0.005. The method use for feature extraction is CC-PEV. The secret message is embedding by nsF5. When the classifier learning with the images with the payload 0.1, 0.05 and natural image and test with the image of the payloads 0.005\end{color}
\begin{figure}[!h]
  \centering
  \includegraphics[width=7cm,height=5cm]{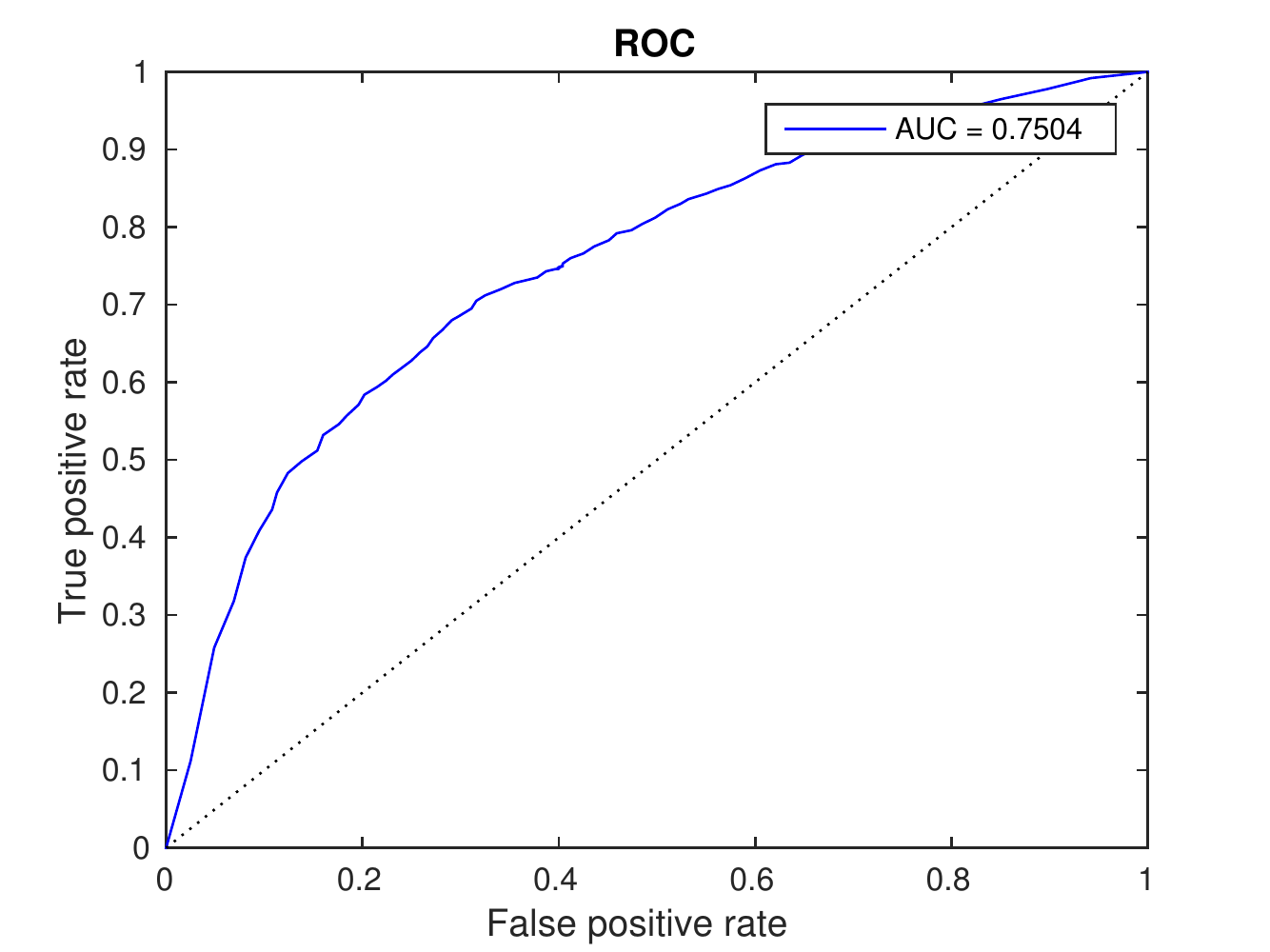}
  \caption{Train with nsF5 with a 0.1 payload and test with J-UNIWARD with a 0.005 payload }
  \label{fig:f5}
 \end{figure}

\section{Conclusion}
\label{sec:conclusion}
This paper has focused on experiments in Kerckhoffs's context: everything about the used steganographic schemes, except the key, are known by steganalysis systems.
Thanks to a large number of experiments, we indeed have shown that even J-UNIWARD can be detected  
but while learning with other steganographic tools, namely HUGO and NSF5.
This is observed even if the objective is to analyse a small payload based steganographic tool. In such a situation, it is sufficient to set a large payload in 
the learning step.
To learn the behavior of a dedicated steganographic scheme, we will consider in a future work to study why less efficient steganographic
tools are more convenient than this dedicated steganographic scheme in the learning process.

%\section*{Acknowledgment}
%The authors acknowledge the University of Franche-Comt\'e, Belfort, France, IUT Belfort et Montb\'eliard,  for providing facilities for conducting experiments and measurements which are reported in this paper and also Diyala University, Diyala, Iraq for providing financial assistance

%\begin{thebibliography}{ref.bib}

%\bibitem{ref:bib}

%\end{thebibliography}
\bibliographystyle{plain}
\bibliography{ref}
\end{document}